\documentclass[12pt,preprint]{aastex}

\begin{document}

\title{Gamma-Ray Bursts in Pulsar Wind Bubbles: Putting the Pieces Together}

\author{Jonathan Granot\altaffilmark{1} and Dafne Guetta\altaffilmark{2}}

\altaffiltext{1}{Institute for Advanced Study, Olden Lane, Princeton, NJ 08540; 
granot@ias.edu}
\altaffiltext{2}{Osservatorio astrofisico di Arcetri, L.E. Fermi 2, Firenze, Italy; 
dafne@arcetri.astro.it}

\begin{abstract}

We present the main observational features expected for Gamma-Ray
Bursts (GRBs) that occur inside pulsar wind bubbles (PWBs). This is
the most natural outcome of supranova model, where initially a
supernova explosion takes place, leaving behind a supra-massive
neutron star, which loses its rotational energy over a time
$t_{\rm sd}$ and collapses to a black hole, triggering a GRB
explosion. We find that the time delay $t_{\rm sd}$ between the
supernova and GRB events is the most important parameter that
determines the behavior of the system. We consider the
afterglow and prompt GRB emission, as well as the direct emission 
from the PWB. The observational
signatures for different ranges in $t_{\rm sd}$ are described and
joined together into one coherent framework. Constraints on the
model are derived for a spherical PWB, from the lack of direct detection
of emission from the PWB together with current afterglow observations.
For very low values of $t_{\rm sd}\lesssim 1\;{\rm hr}$ 
the supranova model reduces to the collapsar model. Values of 
$0.4\;{\rm yr}\lesssim t_{\rm sd}\lesssim 1\;{\rm yr}$ are required 
to produce the iron lines seen in several X-ray afterglows. However,
we find that for a simple spherical model, this implies no
detectable radio afterglow, a small jet break time and
non-relativistic transition time, in disagreement with
observations for some of the GRBs with X-ray lines. These
discrepancies with the observations may be reconciled by resorting
to a non-spherical geometry. 
We find that light element lines, that have been recently detected in a 
few X-ray afterglows, are expected to dominate over iron lines for small 
$t_{\rm sd}$, while for large $t_{\rm sd}$ the situation is reversed.
Finally, we predict that the inverse
Compton upscattering of the PWB photons by the relativistic
electrons of the afterglow (external Compton) should lead to
high energy emission during the early afterglow that might explain
the GeV photons detected by EGRET for a few GRBs, and should be
detectable by future missions such as GLAST.

\end{abstract}


\section{Introduction}
\label{sec:intro}

Despite the large progress in Gamma-Ray Burst (GRB) research over the last several years,
the identity of their progenitors is still one of the most interesting open questions.
Progenitor models of GRBs are divided into two main categories.
The first category involves the merger of a binary system of compact objects, such
as a double neutron star (NS-NS, Eichler et al. 1989), a neutron star and a black hole
(NS-BH, Narayan, Pacy\'nski \& Piran 1992) or a black hole and a Helium star or a white
dwarf (BH-He, BH-WD, Fryer \& Woosley 1998; Fryer, Woosley \& Hartmann 1999). The second
category involves the death of a massive star. It includes the failed supernova (Woosley 1993)
or hypernova (Pacy\'nski 1998) models, where a black hole is created promptly, and a large
accretion rate from a surrounding accretion disk (or torus) feeds a strong relativistic jet
in the polar regions. This type of model is known as the collapsar model. An alternative model
within this second category is the supranova model (Vietri \& Stella 1998), where a massive
star explodes in a supernova and leaves behind a supra-massive neutron star (SMNS) which after
a time delay of $t_{\rm sd}$, loses its rotational energy and collapses to a black hole,
triggering the GRB event. Long GRBs (with a duration $\gtrsim 2\;{\rm s}$) are usually
attributed to the second category of progenitors, while short GRBs are attributed to the first
category. In all the different scenarios mentioned above, the final stage of the process consists
of a newly formed black hole with a large accretion rate from a surrounding torus, and involve a
similar energy budget ($\lesssim 10^{54}\;{\rm ergs}$).

In this work we concentrate on the supranova progenitor model,
focusing on its possible observational signatures. The original
motivation for this model was to provide a relatively baryon clean
environment for the GRB jet. As it turned out, it also seemed to
naturally accommodate the later detection of iron lines in several
X-ray afterglows (Lazzati, Campana, \& Ghisellini 1999; Piro et
al. 2000; Vietri et al. 2001). It has recently been suggested that
the most natural mechanism by which the SMNS can lose its
rotational energy is through a strong pulsar type wind, between
the supernova and the GRB events, which typically creates a pulsar
wind bubble (PWB), also referred to as a plerion (K\"onigl \&
Granot 2002, KG hereafter; Inoue, Guetta \& Pacini 2002).

In an accompanying paper (Guetta \& Granot 2002, GG hereafter) we
study in detail the observational implications of GRBs occurring
inside a PWB. We find that the most important parameter that
determines the behavior of the system is the time delay, $t_{\rm
sd}$, between the supernova and GRB events. The value of $t_{\rm
sd}$ is given by the typical timescale on which the SMNS loses its
rotational energy due to magnetic dipole radiation (see Eq. 2 of
GG) and depends mainly on the polar surface magnetic field
strength of the SMNS, $B_*$ (since its mass, radius and spin
period are constrained to a much smaller range of possible
values). For $B_*\sim 10^{12}-10^{13}\;{\rm G}$, $t_{\rm sd}$
ranges between a few weeks and several years. However, a larger
range in $B_*$, and correspondingly in $t_{\rm sd}$, seems
plausible. We therefore consider $t_{\rm sd}$ as a free parameter.
Another important parameter is the Lorentz factor, $\gamma_w$, of
the pulsar wind, emanating from the SMNS, which is
expected to be in the range $\sim 10^4-10^7$ (KG).

An important difference between our analysis and previous works
(KG; Inoue, Guetta \& Pacini 2002) is that we allow for a proton
component in the pulsar wind, that carries a significant fraction
of its energy. In contrast to the $e^\pm$ component, the internal
energy of the protons in the shocked wind is not radiated away,
and therefore a large fraction of the energy in the pulsar wind
($\sim 10^{53}\;{\rm ergs}$) is always left in the PWB. This
implies that even for a fast cooling PWB, the radius of the wind
termination shock is significantly smaller than the radius of the
supernova remnant (SNR) shell, and that the afterglow shock
typically becomes non-relativistic before it reaches the outer
boundary of the PWB. In the standard model the external medium is
composed of cold protons and electrons (in equal numbers), and has
a density profile that scales with the distance from the source as
$r^{-k}$, where $k=0$ for an ISM and $k=2$ for a stellar wind. In
our scenario, the external medium is made up of hot protons and
cold $e^\pm$ pairs, where there are $\sim 10^3$ times more pairs
than protons. Nevertheless, the protons hold most of the energy in 
the PWB due to their large internal energy, which also dominates the
effective density that  is responsible for the deceleration of the
afterglow shock. The value of $k$ for our model ranges between
$k=0$, that is similar to an ISM, and $k=1$, that is intermediate
between an ISM and a stellar wind.

In this {\it Letter} we relate between the different aspects of
this model, and focus on the observational implications that arise
from different values of $t_{\rm sd}$. We describe the main
results and put them into one coherent picture, while for the
detailed calculations we refer the reader to GG. We show that a
simple spherical model cannot account for the X-ray features
detected in several afterglows, together with the typical
afterglow emission that was observed in the same afterglows. On
the other hand, if the X-ray features turn out not to be real,
then a simple spherical model is compatible with all current
observations, and still holds many advantages compared to other 
progenitor models. It has been pointed out in previous works that
an asymmetry of the remnant is required in order to explain the iron 
lines (Lazzati, Campana, \& Ghisellini 1999; Vietri et al. 2001; KG) 
and its necessity is strengthened by the detailed analysis presented 
in GG, whose main points are reported here. We show that an elongated
PWB can in principle account for the X-ray line together with the
usual afterglow emission. In this paper we also show that this model 
with a modified geometry can in principle account for the iron lines, 
as well as the recent detection of X-ray lines from light elements  
(Reeves et al. 2001; Watson et al. 2002). We find that a small $t_{\rm sd}$ 
favors light element lines, while a large $t_{\rm sd}$ favors iron lines.

In \S \ref{t_sd} we consider the effects of different $t_{\rm sd}$
on the possibility for direct detection of the plerion emission,
the prompt GRB and the afterglow emission, and address the
conditions that are required for the production of iron lines and
not detecting the plerion emission. In \S \ref{X-ray} we discuss the
conditions required for the production of light element lines,
that have been recently observed in a few afterglows, compared to 
iron lines. The high energy emission 
due to the upscattering of the plerion photons by the relativistic
electrons in the afterglow shock (external Compton, EC hereafter)
is discussed in \S \ref{HEE}. In \S \ref{elongated} we outline how 
some of the constraints on the model may be eased if the PWB is
elongated, instead of spherical. Our conclusions are given in \S \ref{conc}.

\section{The Behavior of a Spherical PWB for Different Time Delays}
\label{t_sd}

In this section we go over the main observational signatures of the PWB model,
following the different regimes in $t_{\rm sd}$:

\noindent 1. For extremely small values of $t_{\rm sd}<t_{\rm col}=R_\star/\beta_b c\approx
0.9\,R_{\star,13}\beta_{b,-1}^{-1}\;{\rm hr}$, where $R_\star=10^{13}R_{\star,13}\;{\rm cm}$
is the radius of the progenitor star (before it explodes in a supernova), the stellar 
envelope does not have enough time to increase its radius considerably before the GRB 
goes off, and the supranova model reduces to the collapsar model. In this respect, the 
collapsar model may be seen as a special case of the supranova model. Such low values of 
$t_{\rm sd}$ might be achieved if the SMNS is not rotating uniformly, as differential 
rotation may amplify the magnetic field to very large values, or if the dominant energy 
loss mechanism is gravitational radiation, which can cause significant energy loss on a 
short time scale.

\noindent 2. When $t_{\rm col}<t_{\rm sd}<t_{\rm IS}\sim 16\;{\rm days}$ (e.g. GG) the 
deceleration radius is smaller than the radius for internal shocks. In this case 
the kinetic energy of the GRB ejecta is dissipated through an external shock that 
is driven into the shocked pulsar wind, before internal shocks that result from 
variability within the outflow have time to occur.

\noindent 3. If $t_{\rm IS}<t_{\rm sd}<t_\tau\sim 0.4\;{\rm yr}$, internal
shocks can occur inside the PWB, but the SNR shell is still
optically thick to Thomson scattering, and the radiation from the
plerion, the prompt GRB and the afterglow cannot escape and reach
the observer. If the SNR shell is clumpy (possibly due to the
Rayleigh-Taylor instability, see \S 2 of GG), then the Thomson
optical  depth in the under-dense regions within the SNR shell may
decrease below unity at $t_{\rm sd}$ somewhat smaller than
$t_\tau$, enabling some of the radiation from the plerion to
escape. The only signatures that we expect for this range of
$t_{\rm sd}$ are the neutrino emission due to p-p collisions or
photo-meson interactions, and high energy photons above
$0.5\,(t_{\rm sd}/t_\tau)^{-2}\;$MeV, whose cross section for
scattering on the SNR electrons is reduced due to the
Klein-Nishina effect. Predictions for the neutrino and high energy
photon fluxes from this kind of environment as well as the
mechanisms responsible for this emission will be discussed in
detail in a forthcoming paper (Granot \& Guetta 2002, in
preparation)

\noindent 4. For $t_\tau<t_{\rm sd}<t_{\rm Fe}\sim 1\;{\rm yr}$ the SNR shell has a 
Thomson optical depth smaller than unity, but the optical depth for the iron line 
features is still $\gtrsim 1$ so that detectable X-ray line features, like the iron 
lines observed in several afterglows, can be produced. In this range of $t_{\rm sd}$ 
we expect a very large effective density ($\sim 10^5\;{\rm cm^{-3}}$) and electron 
number density ($\sim 10^3\;{\rm cm^{-3}}$). This effects the afterglow emission in 
a number of different ways: i) The self absorption frequency of the afterglow is 
typically above the radio, implying no detectable radio afterglow, while radio 
afterglows were detected for GRBs 970508, 970828, and 991216, where the iron line 
feature for the latest of these three is the most significant detection to date 
($\sim 4\sigma$, Piro et al. 2000). We also typically 
expect the self absorption frequency of the plerion emission to be above the radio in 
this case, so that the radio emission from the plerion should not be detectable, and 
possibly confused with that of the afterglow. However, for a relatively large iron mass
($\sim 1\;M_\odot$) we can have $t_{\rm Fe}$ as large as $\sim 3-4\;{\rm yr}$, which may 
bring the self absorption frequency of the plerion below the radio band, and thus make 
the radio emission from the plerion detectable (at the level of $\sim 0.1-1\;{\rm mJy}$, 
see Figure 1 of GG). This might provide an alternative explanation for the `enigmatic' 
radio afterglow of GRB 991216 (Frail et al. 2000). ii) A short jet break time $t_j$ and 
a relatively short non-relativistic transition time $t_{\rm NR}$ are implied, as both 
scale linearly with $t_{\rm sd}$ and are in the right range inferred from observations 
for $t_{\rm sd}\sim 30\;{\rm yr}$ (see Eqs. 92, 93 of GG). iii) The electrons are always 
in the fast cooling regime during the entire afterglow.
For $t_{\rm sd}$ in this range the optical emission from the plerion is at the level of
$F_\nu\sim 1\;{\rm\mu Jy}$, for $\gamma_w\lesssim 10^5$.
The X-ray emission from the plerion may become detectable
(i.e. $\gtrsim {\rm a\ few}\; 10^{-14}\;{\rm erg}\;{\rm cm^{-2}}\;{\rm s^{-1}}$)
only for $\gamma_w\lesssim 10^4$ (which is beyond the expected range for $\gamma_w$).

\noindent 5. Finally, for $t_{\rm sd}>t_{\rm Fe}$, we expect no iron lines.
When $t_{\rm sd}$ is between $\sim 2\;{\rm yr}$ and $\sim 20\;{\rm
yr}$ the radio emission of the plerion may be detectable for
$\gamma_w\lesssim 10^5$. The lack of detection of such a radio
emission excludes values of  $t_{\rm sd}$ in this range, if indeed
$\gamma_w\lesssim 10^5$, as is needed to obtain reasonable values
for the break frequencies of the afterglow. For $t_{\rm sd}=t_{\rm
ISM}\sim 38\;{\rm yr}$, the effective density of the PWB is
similar to that of the ISM (i.e. $1\;{\rm cm^{-3}}$), and the
afterglow emission is similar to that of the standard model, where
$k=0$ is similar to an ISM environment, with the exception that in
our model a value of $k=1$, that is intermediate between an ISM
and a stellar wind, is also possible. Larger (smaller) values of
the external density are obtained for smaller (larger) values of
$t_{\rm sd}$. The lack of detection of the EC component in the X-ray 
band ($2-10\;$keV), except perhaps in one afterglow (GRB 000926, 
Harrison et al. 2001) constrains the ratio of the wind termination 
shock radius, $R_s$, and the outer radius of the PWB, $R_b$, to be 
$\lesssim 0.1-0.3$, for $t_{\rm sd}\sim 10-30\;$yr, which is a bit 
hard to obtain with a spherical model (KG).

\section{X-ray Lines: Iron Vs. Light Elements}
\label{X-ray}

Recently, there have been claims for the detection of light element lines 
(Mg, Si, S, Ar, Ca) in the X-ray afterglow of a few GRBs (011211, Reeves et al. 2002;
001025A, 010220, Watson et al. 2002). For GRB 011211, there is no evidence for lines 
from  intermediate mass elements such as Ni, Co or Fe (perhaps only a marginal detection of 
a blueshifted Ni line) and an optical afterglow has been observed (Holland et al. 2002), 
enabling the determination of a spectroscopic redshift, $z=2.141\pm 0.001$ 
(Fruchter et al. 2001; Gladders et al. 2001). For GRBs 001025A and 010220 there is 
no optical afterglow (and therefore no spectroscopic redshift), and there is an 
indication for an over-abundance of Ni or Co (or to our opinion, possibly Fe).
Reeves et al. (2002) estimate the radius of the line producing material from the 
geometrical time delay, $R=t/[(1+z)(1-\cos\theta)]\approx 10^{15}\;$cm, where $\theta$ 
is the angle from which the line photons are emitted (which is identified with the jet 
opening angle, $\theta_j$) and $t$ is the duration of the line emission. However, they 
used $\theta_j=20^\circ$, while there is an indication for a jet break in the optical
light curve at $t_j\approx 1.5-2.7\;$days, which implies $\theta_j\approx 3.4-4.2^\circ$
(Holland et al. 2002). This increases the estimate of the radius $R$ by a factor of 
$[20/(3.4-4.2)]^2\sim 30$. They also estimate the duration of the line emission as 
$10^4\;$s (the time of the observation itself), while we believe that a more reasonable 
estimate is the time after the GRB at which the observation was made. The observation
started $11\;$hr after the burst, while the lines are most prominent during the first 
first $5000\;$s of observation. This would increase the estimate of $R$ by a factor of 
$(11\;{\rm hr})/(10^4\;{\rm s})\approx 4$. Altogether, we obtain 
$R\approx(1.4-2.1)\times 10^{17}\;$cm, which is more than two orders 
of magnitude larger than the estimate of Reeves et al. (2002). The value of $R$ 
may be lower if the ionizing radiation extends out to angles $\theta>\theta_j$.
Therefore, there is no compelling evidence for a small radius of $R\approx 10^{15}\;$cm
and a correspondingly small $t_{\rm sd}\sim R/(0.1c)\sim\;$a few days, just from 
considerations of geometrical time delay. Instead, we obtain $t_{\rm sd}\lesssim 1-2\;$yr.

In order for the light element lines to be stronger than the iron lines 
(or Ni or Co lines for this matter), the ionization parameter should be 
$\xi=4\pi F/n\lesssim 100$ (Lazzati, Ramirez-Ruiz \& Rees 2002), where $F$ is 
the ionizing flux (in the range $1-10\;$keV) and $n$ is the number density of 
the line producing material. We expect a roughly similar ionizing luminosity for 
different GRBs, so that $F\propto R^{-2}$, while the density of the SNR shell
scales as $n\propto 1/R^2\Delta R$, implying $\xi\propto\Delta R$, where $\Delta R$ 
is the width of the SNR shell. We generally expect $\Delta R$ to increase with $R$, 
possibly linearly. Therefore, $\xi$ is expected to increase with $R$ 
and consequently with $t_{\rm sd}$. For this reason we expect the light 
element lines to be more prominent for small values of $t_{\rm sd}$, while Fe
lines should dominate for larger values of $t_{\rm sd}$ (in this case we would not 
expect Ni or Co lines, as the latter would have had enough time to decay into Fe).

\section{External Compton and High Energy Emission}
\label{HEE}

An interesting new ingredient of the PWB model, is that the GRB
and its afterglow occur inside a photon rich plerionic environment.
These photons can be upscattered by the relativistic electrons behind
the afterglow shock, producing a high energy emission (external Compton, EC). 
As has been shown in GG (see Figure 2 therein), for $t_{\rm sd}=t_{\rm ISM}$ 
and $R_s/R_b\lesssim 0.3$, the EC is dominant above $\sim
500\,(t/1\,{\rm hr})^{-1.2}\;{\rm keV}$ (where $t$ is the observed
time after the GRB), while synchrotron is dominant at lower energies. 
This time dependence is valid up to $t\sim
1\;$hr, while for later times the decrease with time is more moderate. 

\begin{figure}
\centering
\noindent
\includegraphics[width=13cm]{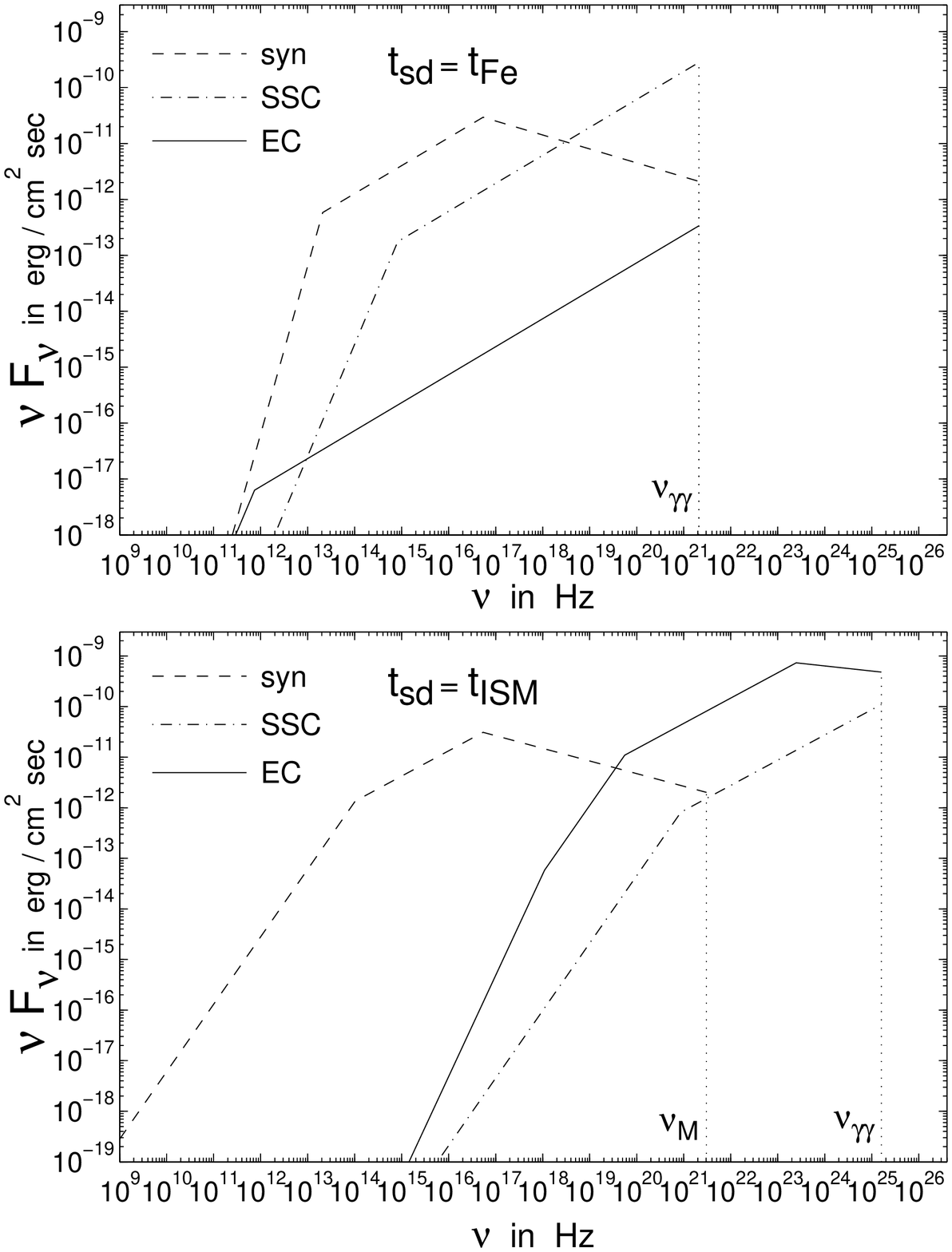}
\caption{\label{fig1}
The afterglow spectrum at one $t=1\;$day after the GRB, for 
$t_{\rm sd}=t_{\rm Fe}\sim 1\;{\rm yr}$ (upper panel) and for 
$t_{\rm sd}=t_{\rm ISM}\sim 38\;{\rm yr}$ (lower panel), calculated 
for the fiducial parameters of Guetta \& Granot (2002). 
Dotted vertical lines indicate $\nu_M$ where the upper cutoff for the 
synchrotron emission is located (e.g. GG), and $\nu_{\gamma\gamma}$ where
the upper cutoff of the SSC and EC (due to pair opacity) is located.
}
\end{figure}

Figure \ref{fig1} shows the afterglow spectrum at $t=1\;$ day,
for $t_{\rm sd}=t_{\rm Fe}$ and $t_{\rm ISM}$, where for clarity,
the synchrotron, synchrotron self-Compton (SSC) and EC components 
are shown separately. It can be seen that the EC component becomes
more important for larger $t_{\rm sd}$. We expect an upper cutoff 
due to opacity to pair production with the photons of the plerion at 
$h\nu_{\gamma\gamma}\sim 100(t_{\rm sd}/t_{\rm ISM})^2\;$GeV. 
This latter upper cutoff moves down to a lower energy for smaller values of 
$t_{\rm sd}$, and is $\sim 100\;$MeV for $t_{\rm sd}=1\;{\rm yr}\sim t_{\rm Fe}$,
as can be seen in Figure \ref{fig1}. For afterglows with X-ray 
line features we expect no high energy emission above this limit. 

For $t_{\rm sd}\sim t_{\rm ISM}$, the EC component dominates the 
early afterglow ($t\lesssim 100\;$s) emission above $\sim 100\;$MeV. 
At early times, the afterglow radius is relatively small and we 
expect the ratio, $X$, of energies in the EC and
synchrotron components to be roughly constant in time, so that the
peak of the $\nu F_\nu$ EC spectrum has a temporal scaling similar
to that of the synchrotron component (i.e. $\propto t^{-1}$, see
GG). We expect $\nu F_\nu$ to decay
very slowly with time, as $t^{-1/4}$, for $\nu<\nu_m^{EC}$, and
decay approximately linearly with time ($\propto t^{-1-3(s-2)/4}$)
for $\nu>\nu_m^{EC}$. The temporal decay becomes steeper than these
scalings as the afterglow radius increases and $X$
begins to decrease with time. The EC emission can account for the
high energy emission detected by EGRET for GRB 940217 (Hurley et
al. 1994), and is consistent with the flux level and relatively
moderate time decay observed in this case.

A different interpretation for the high energy emission discussed
above was recently suggested by Wang, Dai \& Lu (2002), in a
similar context of the supranova model, where the GRB occurs
inside a plerionic environment. However, their results imply that
the typical synchrotron frequency is $h\nu_m\sim$ a few keV after
one day, which is inconsistent with afterglow observations (unless
GRBs with delayed high energy emission constitute a different
class of GRBs, with a very different afterglow emission).

\section{An Elongated Geometry}
\label{elongated}

For a spherical PWB, the X-ray features observed in several afterglows 
cannot be reconciled with the conventional afterglow emission observed 
in the same afterglows. However, this discrepancy may be reconciled if 
the PWB is elongated along its rotational axis, so that the polar radius is
much larger than the equatorial radius. One may naturally expect
such a geometry for a number of different reasons (KG; GG). 
In this case the iron lines can be emitted by clumpy SNR material at small 
radii, near the equator, while the afterglow emission originates from along
the polar direction, where the GRB outflow is expected to
propagate, and may reach a considerably larger radius. In this
picture, the effective density within the PWB is relatively small,
close to that of a sphere with the polar radius. This helps
reproduce the typical afterglow emission, and avoid direct detection 
of the plerion emission in the radio. Another advantage of an
elongated geometry is the suppression of the EC component in the
X-ray afterglow, that results since a ratio of $R_s/R_b\lesssim
0.1$ is naturally expected in this case.

\section{Conclusions}
\label{conc}

In this work we have presented the main observational implications for GRBs 
that occur inside pulsar wind bubbles (PWBs), as expected in the supranova
model. We have examined the relations between the different observations
and put the different ingredients of the model into one
coherent framework.

We find that a simple spherical model cannot produce the iron line
features observed in several afterglows together with the other,
more conventional, features of the afterglow emission from these
bursts. However, if the iron lines are not real, then a simple
spherical model can explain all other observations for $t_{\rm
sd}\gtrsim 20\;$yr. The latter is required in order to explain
typical afterglow observations and the lack of direct detection 
of the plerion emission in the radio during the afterglow.

If the iron line detections are real, then in the context of the
PWB model, this requires deviations from a simple spherical
geometry. The most straightforward variation of the simple model
is a PWB that is elongated along its polar axis. Such a geometry
may arise naturally within the context of this model (KG; GG).

With an elongated geometry, the PWB model can account for all the
observed features in the afterglow, and it offers a number of
advantages in comparison to other models: 
i) It provides a relatively baryon clean environment
for the GRB jet, which is required in order to produce a highly
relativistic outflow. This arises as the initial supernova expels
most of the stellar envelope to a large distance from the site of
the GRB, and the strong pulsar wind effectively sweeps up the remaining 
baryonic matter. ii) An important advantage of this
model is that it can naturally explain the large values of
$\epsilon_B$ and $\epsilon_e$ that are inferred from fits to
afterglow data (KG), thanks to the large magnetic fields in the
PWB and the large relative number of electron-positron pairs. This
is in contrast with standard environment that is usually assumed
to be either an ISM or the stellar wind of a massive progenitor,
that consists of protons and electrons in equal numbers. In this 
case, the pre-existing magnetic field, that is amplified due to the 
compression of the fluid in the shock, is too small to explain the values 
inferred from afterglow observations, and further magnetic 
field amplification or generation at the shock is required. 
iii) All the detections of GRB afterglows to date 
are for the long duration sub-class of GRBs (with a duration $\gtrsim
2\;{\rm s}$), that are believed to arise from a massive star
progenitor, which according to the collapsar model should imply a
stellar wind environment ($k=2$). However, a homogeneous external
medium ($k=0$) provides a better fit to the observational data for
most GRB afterglows. This apparent contradiction is naturally
explained in the context of the PWB model, where $k$ ranges
between $0$ and $1$, while we still have a massive star
progenitor. iv) Another advantage of the PWB model is its
capability of explaining the high energy emission observed in some
GRBs (Schneid et al. 1992; Sommer et al. 1994; Hurley et al.
19994; Schneid et al. 1995). We find that the high energy emission
during the early afterglow at photon energies $\gtrsim 100\;$keV
is dominated by the EC component. We predict that such a high
energy emission may be detected in a large fraction of GRBs with
the upcoming mission GLAST. However, we find an upper cutoff at a
photon energy of $\sim 100\,(t_{\rm sd}/1\;{\rm yr})^2\;$MeV, due
to opacity to pair production with the photons of the PWB. This
implies no high energy emission above $\sim 100\;$MeV for
afterglows with X-ray line features, but allows photons up to an
energy of $\sim 100\;$GeV for afterglows with an external density
typical of the ISM ($t_{\rm sd}\sim 38\;$yr).

\acknowledgements

We thank Arieh K\"onigl for many helpful discussions.
This research was supported by the Institute for Advanced Study, 
funds for natural sciences (JG).
We thank the Einstein Center at the Weizmann Institute of Science for the
hospitality and for the pleasant working atmosphere.
DG thanks the Institute for Advanced Study, where most of this research 
was carried out, for the hospitality and the nice working atmosphere.


\end{document}